\documentclass[10pt,conference,anonymous]{IEEEtran}
\IEEEoverridecommandlockouts

\usepackage{mdframed}
\usepackage{tcolorbox}
\usepackage{listings}
\usepackage[T1]{fontenc}
\usepackage[utf8]{inputenc}
\usepackage{float}
\usepackage{multirow}
\usepackage{amsthm}
\usepackage{graphicx}
\usepackage{changepage}
\usepackage{url}
\usepackage{amssymb}
\PassOptionsToPackage{normalem}{ulem}
\usepackage{ulem}
\usepackage{array}
\usepackage{booktabs}
\usepackage{amsmath}
\usepackage{stackrel}
\usepackage{babel}
\usepackage{amsmath}
\usepackage{amsthm}
\usepackage{pgfplots}
\usepackage{pgf-pie}

\usepackage{balance}
\usepackage{hyperref}
\usepackage{xcolor}

\usepackage{pgfplots}
\usepackage{pgfplotstable}
\pgfplotsset{compat=1.7}
\usepackage{tikz}
\usepackage{framed}
\usepackage{multirow}
\usepackage{enumerate}
\usepackage{graphicx}
\usepackage{threeparttable}

\hypersetup{
  colorlinks   = True, 
  urlcolor     = blue, 
  linkcolor    = blue, 
  citecolor    = blue 
}

\begin{document}

\title{Security Defect Detection via Code Review: A Study of the OpenStack and Qt Communities}

\author{
    \IEEEauthorblockN{Jiaxin Yu$^{1,2}$, Liming Fu$^{1,2}$, Peng Liang$^{1,2*}$\thanks{\indent This work is funded by the NSFC with Grant No. 62172311 and the Special Fund of Hubei Luojia Laboratory. Amjed Tahir is supported by a MU SREF grant.}, Amjed Tahir$^{3}$, Mojtaba Shahin$^{4}$}
    \IEEEauthorblockA{$^1$ School of Computer Science, Wuhan University, Wuhan, China}
    \IEEEauthorblockA{$^2$ Hubei Luojia Laboratory, Wuhan, China}
    \IEEEauthorblockA{$^3$ School of Mathematical and Computational Sciences, Massey University, Palmerston North, New Zealand}
    \IEEEauthorblockA{$^4$ School of Computing Technologies, RMIT University, Melbourne, Australia}
    \IEEEauthorblockA{\{jiaxinyu, limingfu, liangp\}@whu.edu.cn, a.tahir@massey.ac.nz, mojtaba.shahin@rmit.edu.au}
}


\IEEEoverridecommandlockouts
\IEEEpubid{\makebox[\columnwidth]{978-1-6654-5223-6/23/\$31.00~\copyright2023 IEEE\hfill} \hspace{\columnsep}\makebox[\columnwidth]{ }}
\maketitle 
\IEEEpubidadjcol
\begin{abstract}
\textit{Background}: Despite the widespread use of automated security defect detection tools, software projects still contain many security defects that could result in serious damage. Such tools are largely context-insensitive and may not cover all possible scenarios in testing potential issues, which makes them susceptible to missing complex security defects. Hence, thorough detection entails a synergistic cooperation between these tools and human-intensive detection techniques, including code review. Code review is widely recognized as a crucial and effective practice for identifying security defects. 
\textit{Aim}: This work aims to empirically investigate security defect detection through code review.
\textit{Method}: To this end, we conducted an empirical study by analyzing code review comments derived from four projects in the OpenStack and Qt communities. Through manually checking 20,995 review comments obtained by keyword-based search, we identified 614 comments as security-related. 
\textit{Results}: Our results show that (1) security defects are not prevalently discussed in code review, (2) more than half of the reviewers provided explicit fixing strategies/solutions to help developers fix security defects, (3) developers tend to follow reviewers' suggestions and action the changes, 
(4) \textit{Not worth fixing the defect now} and \textit{Disagreement between the developer and the reviewer} are the main causes for not resolving security defects. 
\textit{Conclusions}: Our research results demonstrate that (1) software security practices should combine manual code review with automated detection tools, achieving a more comprehensive coverage to identifying and addressing security defects, and (2) promoting appropriate standardization of practitioners' behaviors during code review remains necessary for enhancing software security.
\end{abstract}

\begin{IEEEkeywords}
Code Review, Security Defect, OpenStack, Qt, Empirical Study
\end{IEEEkeywords}


\maketitle

\section{Introduction}

Security defects can have serious consequences, such as data breaches, intellectual property theft and disruption of services~\cite{telang2007empirical,cavusoglu2004effect}. Numerous studies have emphasized the significance of keeping software under control to reduce the risk of exploitation~\cite{iannone2022secret, mcgraw2013software, planning2002economic}. Nevertheless, the practice of leaving a large number of security defects unaddressed in the production environment for extended periods of time and only patching them after they have been released publicly ~\cite{alfadel2023empirical}, has a negative impact on software quality and leads to increased maintenance costs. Therefore, effectively minimizing the financial and reputational costs of security incidents by detecting security defects as early as possible remains the major focus for the stakeholders involved in software production. 

Many organizations are shifting security practices to earlier stages of software development, hoping to address security concerns before they become more difficult and expensive to fix~\cite{gitlab2022url}. Under this circumstance, code review is proven to be an effective method to identify and locate security defects early~\cite{bosu2014identifying, thompson2017large}. Code review is a valuable practice of systematically and internally examining revisions before code is released to production to detect defects and ensure quality. Code review is one of the most important practices of modern software development~\cite{bosu2016process}. Compared with security defect detection tools, code review participants are mostly project members who can take full account of the code context~\cite{mcconnell2004code}; thus, they are in a position to identify security defects effectively.

Several studies have focused on security defects detection in code review (e.g.,~\cite{edmundson2013empirical, paul2021improving, di2016security, alfadel4161317qualitative, bosu2014identifying}). Bosu \textit{et al.} investigated the distribution and characteristics of security defects identified by code reviewers~\cite{bosu2014identifying}, while Paul \textit{et al.} focused on the security defects that were missed during code review~\cite{di2016security}. However, most of the research mainly concentrated on the identification of security defects, rather than delving into their resolution procedures. Specifically, little is known about the actions taken by practitioners and the challenges they face when resolving identified security defects in code review. Exploring these aspects could help increase the fixing rate of identified security defects during code review.


To this end, this work \textbf{aims} to explore the resolution of security defects through the means of code review,
thus contributing to develop a more comprehensive body of knowledge on security defect detection via code review. We first collected 432,585 review comments from four active projects of two widely known communities: OpenStack (Nova and Neutron) and Qt (Qt Base and Qt Creator). After a keyword-based search on these review comments, we manually analyzed 20,995 potential security-related comments, resulting in 614 comments that actually identified security defects. We then studied the types of security defects identified, how the practitioners treat the identified defects, and why some of them are finally unresolved in code review. 

Our \textbf{findings} show that: (1) security defects are not widely identified in code review; (2) when faced with security defects, most reviewers express their opinions on fixing them and provide specific solutions, which are generally agreed and adopted by developers; 
(3) \textit{Disagreement between the developer and reviewer} and \textit{Not worth fixing the defect now} are the most frequent causes of not resolving security defects. 


The \textbf{contributions} of this work are: (1) We highlight the importance of manual and context-sensitive security review of code, which may reveal security defects undetected by automated tools. (2) We complement the datasets of previous works on the types of security defects identified during code review. (3) We provide the best practices for practitioners' behaviour in modern code review for security defects detection.


\section{Related Work}
\label{sec:related-work}

\subsection{Security Defect Detection}

A body of research has focused on the current status of security defect detection across software ecosystems. Alfadel \textit{et al.} discussed vulnerabilities propagation, discovery, and fixes in Python ecosystem~\cite{alfadel2023empirical}. It was found that most exposed security defects were not being fixed in a timely manner. A similar study of \texttt{npm} packages demonstrated that delays in fixing security defects were often caused by the fact that the fix was bundled with other features and did not receive the necessary prioritization~\cite{chinthanet2021lags}. Lin \textit{et al.} investigated the security defect management in Debian and Fedora ecosystems~\cite{lin2023vulnerability}, and found that over 50\% of security defects fixes in Linux distributions can be integrated within one week. Our work differs from the aforementioned studies in that the security defects discussed in these works are publicly disclosed, while in our work we focused on security defects that practitioners may notice during their daily coding activities (but may not have been already disclosed).

Security defects can be detected through automated approaches or manually. Tudela \textit{et al.} utilized hybrid analysis to detect the OWASP Top Ten security vulnerabilities and discussed the performance of different tool combinations~\cite{mateo2020combining}. Singh \textit{et al.} compared the difference in automated (belong to DAST) and manual approaches for penetration testing, indicating that humans can locate security defects missed by automated scanners~\cite{singh2020automated}. Osterweil \textit{et al.} formulated a framework using IAST to improve human-intensive approaches in security defect detection and proved its effectiveness~\cite{osterweil2017comprehensive}.
Inspired by the above-mentioned studies, we were motivated to explore an effective human-intensive practice for detecting security defects, i.e., code review, and pave the way for further integrating automated tools into the code review process.

\subsection{Security Defect Detection in Code Review}

Several studies have studied security defect detection in code review. For example, 
di Biase \textit{et al.} explored the value of modern code review for system security and investigated the factors that can affect security testing based on the Chromium project ~\cite{di2016security}. Thompson \textit{et al.} conducted a large-scale analysis of the dataset obtained from GitHub~\cite{thompson2017large} and reaffirmed the crucial relationship between code review coverage and software security.

There is a growing interest in improving the effectiveness of security code review. Paul \textit{et al.} analyzed 18 attributes of a code review to explore factors that influence the identification of security defects, in order to pinpoint areas of concern and provide targeted measures~\cite{paul2021security}. Braz \textit{et al.} analyzed the impact of two external assistance measures on the identification of security defects~\cite{braz2022less} and found that explicitly requiring practitioners to concentrate on security can greatly increase the probability of finding security defects, while the further provision of security checklists did not show better results.

Some studies qualitatively analyzed the implementation of security defect detection in code review. Alfadel \textit{et al.} investigated security-related reviews in \texttt{npm} packages \cite{alfadel4161317qualitative} to analyze the proportion, types, and solutions of identified security defects in these reviews. In comparison, we targeted different data sources and provided a more in-depth analysis, which includes the causes for not resolving security defects and the actions of developers and reviewers when facing security defects in code review; therefore providing a holistic understanding of the current status of security code review. Motivated by these related works, we aim to bridge the knowledge gap with a view to inspire new research directions and enhance the effectiveness of detecting security defects.

\section{Methodology}
\label{sec:method}

\subsection{Research Questions}
The goal of this study is to examine the implementation of security defect detection in code review. Specially, we analyzed review comments to investigate how security defects are identified, discussed, and resolved by reviewers and developers. 
To achieve this goal, we formulated the following Research Questions (RQs):  

\noindent\textbf{RQ1:} \textit{What types of security defects are identified in code reviews}? \\Previous studies have explored the distribution of security defects found in code reviews~\cite{bosu2014identifying, di2016security, alfadel4161317qualitative, bosu2013peer}. However, those studies have largely focused on specific systems and the types of security defects may vary in different systems, warranting additional research encompassing diverse projects to establish more general findings~\cite{di2016security}. Driven by this, RQ1 investigates the frequency of each security defect type within the OpenStack and Qt communities, aiming to complement the findings from existing studies. 


\noindent\textbf{RQ2:} \textit{How do developers and reviewers treat security defects identified in code reviews?} \\ Given that strict reviewing criteria were mostly abandoned in modern code review~\cite{sadowski2018modern}, it is necessary to establish a good understanding of the current practices employed by practitioners and how they influence the quality of security code review, so as to capture the undesirable behaviors and formulate corresponding suggestions for best practices. This RQ aims to explore concrete actions of developers and reviewers after security defects were identified. Answering this RQ helps to better understand the resolution process and the extent to which manual security defect detection is implemented in code review. In addition, the common solutions of each security defect type extracted from the changed source code can be used to support developers in addressing security defects in the future. This RQ is further decomposed into four sub-RQs:


\textbf{RQ2.1:} \textit{What actions do reviewers suggest to resolve security defects?}

\textbf{RQ2.2:} \textit{What actions do developers take to resolve security defects?}

\textbf{RQ2.3:} \textit{What is the relationship between the actions suggested by reviewers and those taken by developers?}

\textbf{RQ2.4:} \textit{What are the common solutions to each security defect type identified in code reviews?}






\noindent\textbf{RQ3:} \textit{What are the causes for developers not resolving the identified security defects?} \\ In some cases, security defects are identified by reviewers but not ultimately resolved by developers. However, little research has been conducted to understand the reasons behind these cases, which could shed light on potential obstacles developers encounter and help in facilitating the resolution of identified security defects. As a result, RQ3 explores potential causes of why some defects are not fixed, with the objective of filling this gap and providing valuable insights.


\subsection{Data Collection}
\begin{figure*}[htbp]
  \centering
  \includegraphics[width=\linewidth]{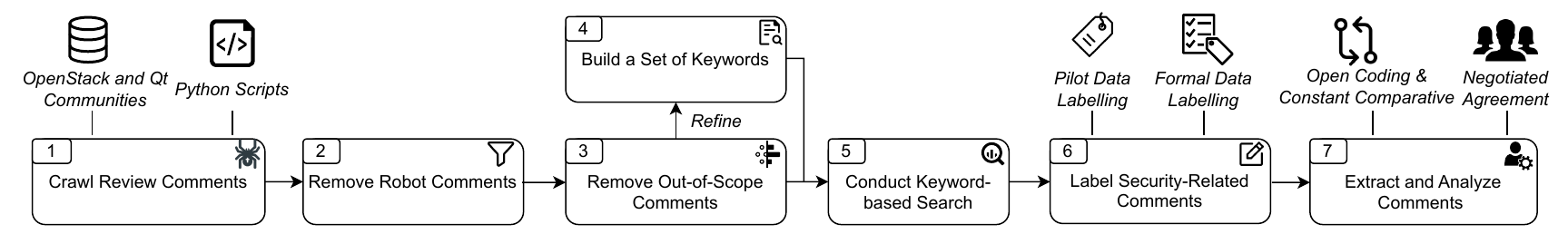}
  \caption{ An overview of our data acquisition, processing and analysis process}
  \label{fig:overview}
\end{figure*}

The data collection, labelling, extraction, and analysis process is described below (an overview is shown in Fig.~\ref{fig:overview}).
\subsubsection{Projects Selection}

This study analyzes security defects in code reviews collected from four projects of two communities: Nova\footnote{\url{https://github.com/openstack/nova}} and Neutron\footnote{\url{https://github.com/openstack/neutron}} from OpenStack\footnote{\url{https://www.openstack.org/}}, and Qt Base\footnote{\url{https://github.com/qt/qtbase}} and Qt Creator\footnote{\url{https://github.com/qt-creator/qt-creator}} from Qt\footnote{\url{https://www.qt.io/}}. These two communities are selected based on the following two criteria~\cite{mcintosh2014impact}: 1) \textit{Reviewing Policy} - the community has established a strong review process, and 2) \textit{Traceability} - the review process of the community should be traceable. 

OpenStack is a platform that builds and manages public or private cloud, with a set of projects responsible for processing different core cloud computing services. 
Qt is a cross-platform application for creating GUI applications. We deemed these two communities to be appropriate for our study as they have a large number of code reviews, which are performed using a traceable code review tool - Gerrit\footnote{\url{https://www.gerritcodereview.com/}}. Gerrit offers on-demand tracking of the review process~\cite{thongtanunam2017review}. The projects from the two communities have been widely used in previous code review studies (e.g., \cite{spadini2018testing,hamasaki2013does,han2022code,fu2022potential}). Similar to Hirao \textit{et al.}~\cite{hirao2020code}, we selected two active projects from OpenStack (i.e., Nova and Neutron) and Qt (i.e., Qt Base and Qt Creator), which have the highest number of patches.

\subsubsection{Review Comments Collection}
\label{sec:total_data_collection}

Using the RESTful API provided by Gerrit, we obtained a total of 432,585 review comments from the four projects (166,237 review comments from OpenStack and 266,348 from Qt) spanning from January 2017 to June 2022, the time when we started this work. Considering that our study aims to analyze the practices of developers and reviewers when dealing with security defects, any comments made by bots should be excluded. Hence, we filtered out the review comments of which the author is a bot account (i.e., ``\textit{Zuul}'' in OpenStack and ``\textit{Qt Sanity Bot}'' in Qt). We also removed review comments in files that do not correspond to any programming language or are clearly outside the scope of code review, by checking the filename extension (e.g., ``\texttt{.orig}'' and ``\texttt{.svg}'').

\subsubsection{Potential Security-related Comments Collection}
We employed a keyword-based search approach to identify security-related review comments 
We adopted the keyword set proposed in Paul \textit{et al.}'s work~\cite{paul2021security}, as it is considered the most comprehensive keyword set in previous research, with the largest number of types and keywords. 
The set includes 103 keywords, which were classified into 11 security defect types and an extra \textit{Common Keywords} type, with each security defect type containing Common Weakness Enumerations (CWEs)~\cite{cwe2022url} to clarify its definition. After thoroughly analyzing the keyword set proposed by Paul \textit{et al.}~\cite{paul2021security}, we made the following adjustments to the set:

First, we adapted parts of the types of security defect and corresponding keywords. For example, we split \textit{Denial of
Service (DoS) } from the \textit{Denial of Service (DoS) / Crash} type defined in Paul \textit{et al.}'s work~\cite{paul2021security}, since we considered \textit{DoS} as one clear security defect type based on its definition in CWEs. The keywords relevant to \textit{DoS} were also separated and reclassified into the new \textit{DoS} type.

Second, we collected differentiated keywords and security defect types from previous studies~\cite{di2016security,bosu2013peer} and extended the keyword set obtained from the last step. One additional security defect type was added (i.e., the \textit{Command Injection} type~\cite{bosu2013peer}). Moreover, another one additional security defect type was created since part of keywords from~\cite{di2016security} could not be mapped into the existing keyword set (i.e., \textit{Use After Free} was created to include ``\textit{use-after-free}'' and ``\textit{dynamic}'' based on the definition of CWEs). 19 differentiated keywords collected from previous studies were assigned to specific types (including \textit{Common Keywords}) according to their meanings, (e.g., adding ``\textit{crypto}'' to the \textit{Encrypt} type).

After that, the initial keyword set of our study was formulated and presented in Table~\ref{tab:keywords}. We ultimately obtained 122 keywords, which were categorized into 15 security defect types and the \textit{Common Keywords} type. To explicitly illustrate our adjustments, the sources of each type are presented, and newly added keywords compared to the keywords from Paul et al.'s work~\cite{paul2021security} are emphasized in italics.

Given that the effectiveness of the keyword-based approach heavily depends on the set of keywords used, we followed the approach proposed by Bosu \textit{et al.}~\cite{bosu2014identifying} to refine the initial set of keywords, which includes the following steps:
\begin{enumerate}
    \item build a corpus by searching for review comments that contain at least one keyword of our initial set of keywords (e.g., ``\textit{racy}'', ``\textit{overflow}'') in the review comments collected in Section~\ref{sec:total_data_collection}.
    \item perform tokenization to each document on the corpus. Considering code snippets contained in review comments, we also applied the identifier splitting rules in this progress (e.g., ``\textit{FlavorImageConflict}'' becomes ``\textit{Flavor Image Conflict}'', \textit{security\_group} becomes ``\textit{security group}'').
    \item remove stopwords, punctuations, and numbers from the corpus and convert all tokens into lowercase.
    \item use \texttt{SnowballStemmer} from the NLTK toolkit~\cite{bird2009natural} to obtain the stem of each token (e.g., ``\textit{merged}'', ``\textit{merging}'', and ``\textit{merges}'' have the same token ``\textit{merg}'').
    \item create a Document-Term matrix ~\cite{tan2016introduction} from the corpus and identify the additional words that frequently co-occur with each of our initial keywords (co-occurrence probability of 0.05 in the same document, as also utilized in~\cite{bosu2014identifying}).
    \item manually analyze the additional words to determine whether to include them into the initial keyword set.
\end{enumerate}

No additional words were found that co-occurred with any one of the initial keywords. Therefore, we were of the opinion that the present keyword set is adequate for supporting keywords-based search and filtering. After that, a script was developed to search for code review comments that contain at least one of the keywords identified in Table \ref{tab:keywords}. All these steps led to 20,995 review comments from the four projects, which is called \textbf{potential security-related review comments}.

\begin{table*}[htbp]
\centering
\scriptsize
  \caption{Keywords to mine code reviews that identify security defects}
  \label{tab:keywords}
  \begin{threeparttable}
  \begin{tabular}{ |p{3cm}|p{2cm}|p{2.3cm}|p{8.5cm}| } 
    \hline
    \textbf{Security Defect Type}   &\textbf{Source}                            &\textbf{CWE ID}                   &\textbf{Keywords\tnote{*}}                                                                               \\\hline
    Race Condition                  &~\cite{paul2021security}                   &362-364, 366-368                  & race, racy                                                                          \\\hline
    Crash                           &Adapted from~\cite{paul2021security}       &248, 754, 755                     & {crash, \textit{exception}}                                                                     \\\hline
    Resource Leak                   &Adapted from~\cite{paul2021security}      &401, 404                          & {leak}                                                                                 \\\hline
    Integer Overflow                &~\cite{paul2021security}                  &190, 191, 680                     & {integer, overflow, signedness, widthness, underflow}                                  \\\hline
    Improper Access                 &~\cite{paul2021security}                  &22, 264, 269, 276, 281-290      & {improper, unauthenticated, gain access, permission, hijack, authenticate, privilege, forensic, hacker, root, \textit{URL, form, field, sensitive}}                                                                                                                  \\\hline
    Buffer Overflow                 &~\cite{paul2021security}                  &120-127                         & {buffer, overflow, stack, strcpy, strcat, strtok, gets, makepath, splitpath, heap, strlen, \textit{out of memory}}                                                                                                                                                    \\\hline
    Denial of Service (DoS)         &Adapted from~\cite{paul2021security}      &400, 402, 403, 405-406          & {denial service, dos, ddos}                                                            \\\hline
    Deadlock                        &~\cite{paul2021security}                  &833                               & {deadlock}                                                                             \\\hline
    Encryption                      &~\cite{paul2021security}                  &310, 311, 320-327               & {encrypt, decrypt, password, cipher, trust, checksum, nonce, salt, \textit{crypto, mismatch}}   \\\hline 
    Cross Site Scripting (XSS)      &~\cite{paul2021security}                  &79-87                           & {cross site, CSS, XSS, malform, \textit{htmlspecialchar}}                               \\\hline
    Use After Free                  &Created in this work                             &416                               & {\textit{use-after-free, dynamic}}                                                                         \\\hline
    Command Injection               &~\cite{bosu2013peer}                      &77-78, 88                       & {\textit{command, exec}}                                                             \\\hline
    Cross Site Request Forgery      &~\cite{paul2021security}                  &352                               & {cross site, request forgery, CSRF, XSRF, forged, \textit{cookie, xhttp}}                       \\\hline
    Format String                   &~\cite{paul2021security}                  &134                               & {format, string, printf, scanf, \textit{sanitize}}                                              \\\hline
    SQL Injection                   &~\cite{paul2021security}                  &89                                & {SQL, SQLI, injection, \textit{ondelete}}                                                       \\\hline
    Common Keywords                 &~\cite{paul2021security}                  &-                                 & {security, vulnerability, vulnerable, hole, exploit, attack, bypass, backdoor, threat, expose, breach, violate, fatal, blacklist, overrun, insecure, scare, scary, conflict, trojan, firewall, spyware, adware, virus, ransom, malware, malicious, risk, dangling, unsafe, steal, worm, phishing, cve, cwe, collusion, covert, mitm, sniffer, quarantine, scam, spam, spoof, tamper, zombie, \textit{cast, xml}}                \\\hline 
\end{tabular}
\begin{tablenotes}
    \footnotesize
    \item[*] Most of the keywords in this list are adopted from the prior study of Paul \textit{et al.}~\cite{paul2021security}. The keywords in italic are our additions to this list.
\end{tablenotes}
\end{threeparttable}
\end{table*}


\subsection{Manual Labelling}
\label{sec:data_label}
The 20,995 potential security-related review comments obtained from the previous step may contain many false positives. Hence, we manually inspected the content of these comments, their corresponding discussions, and related source code to determine and label whether they are actually security-related. We defined the labelling criteria, i.e., the review comment should be clearly related to security and meet the definition of one of the CWEs~\cite{cwe2022url} presented in Table \ref{tab:keywords}. Aimed at ensuring consistency and improving inter-rater reliability, a pilot labelling was independently conducted by the first and second authors on 200 potential security-related comments randomly selected from the Nova project. The labelling results were compared and the level of agreement between the two authors was measured using Cohen's Kappa coefficient test~\cite{cohen1960coefficient}. For review comments in which the judgements of two raters differ, they were reviewed, evaluated, and discussed with the third author until a consensus was reached. The calculated Cohen's Kappa coefficient is 0.87, thus indicating that the two authors reached a high level of agreement. The first author proceeded to label all the remaining potential security-related comments, and the review comments that the first author was unsure were discussed with the second author to reach a consensus. This process led to the identification of a total of 614 \textbf{security-related review comments} for further analysis and the distribution of data points across the four projects is presented in Table~\ref{tab:project_distribution}:

\begin{table}[htbp]
    \centering
    \scriptsize
    \caption{Keywords to mine code reviews that identify security defects}
    \label{tab:project_distribution}
    \begin{tabular}{|c|c|c|c|}
    \hline
         \textbf{Project}      & \textbf{Comments}    & \textbf{Potential Security-related}      & \textbf{Security-related}  \\\hline
         Neutron      & 56,846              & 3,289                            & 88               \\\hline
         Nova         & 109,391             & 7,677                            & 192              \\\hline
         Qt Base      & 170,820             & 7,677                            & 241              \\\hline
         Qt Creator   & 95,528              & 2,352                            & 93               \\\hline
         \textbf{Total}        & 432,585             & 20,995                           & 614              \\\hline
    \end{tabular}
\end{table}

\subsection{Data Extraction and Analysis}
A set of data items (see Table \ref{tab:Dataitems}) was formulated and extracted from the contextual information of each of the 614 security-related comments, including their corresponding discussion thread and source code, to answer our RQs.
\subsubsection{RQ1}
We classified 614 security-related review comments into 15 security defect types predefined in Table \ref{tab:keywords}. Based on this table, for each review comment, we identified the CWE corresponding to the issue described in the comment, and categorized the comment under the security defect type to which that CWE belongs. As shown in the example below, the reviewer pointed out that the calculation of \verb|pos+n| may overflow and lead to undefined behavior, which is consistent with the description of CWE-109, that is ``\textit{The software performs a calculation that can produce an integer overflow or wraparound, when the logic assumes that the resulting value will always be larger than the original value}'', hence is labelled as \textit{Integer Overflow}.

\begin{lstlisting}[label={lst:sample},breaklines=true,breakatwhitespace=true,captionpos=b,basicstyle=\footnotesize\ttfamily,escapeinside=||,frame=single]
|\textbf{Link:}| |\url{http://alturl.com/td7ej}|  
|\textbf{Project:}| Qt Base      
|\textbf{Type:}| Integer overflow      
|\textbf{Reviewer: This is UB when pos + n overflows...}| 
|\textbf{Developer:}| Done
\end{lstlisting}





\subsubsection{RQ2}
\noindent
We categorized the actions suggested by reviewers into three categories with reference to what was formulated by Tahir \textit{et al.} in~\cite{tahir2018can,tahir2020large}.
\begin{enumerate}
    \item\textbf{Fix}: recommend fixing the security defect.
    \item\textbf{Capture}: detect the security defect, but do not provide any further guidance.
    \item\textbf{Ignore}:  recommend ignoring the security defect.
\end{enumerate}

Confronted with review comments posted by reviewers, there are three possible behaviors for developers:
\begin{enumerate}
    \item\textbf{Resolve}: The developer resolved the security defect identified by the reviewer.
    \item\textbf{Not resolve}: The developer ignored the security defect identified by the reviewer.
    \item\textbf{Unknown}: We are unable to determine the behavior of the developer.
\end{enumerate}

We defined \textit{Unknown} to describe the case that the developer responds to the reviewer with a promise to fix the security defect in the future, but we could not obtain specific resolution evidence from the source code due to the overwhelming amount of manual inspection of unlimited commits. An example of such a case is shown below:
\begin{lstlisting}[label={lst:sample},breaklines=true,breakatwhitespace=true,captionpos=b,basicstyle=\footnotesize\ttfamily,escapeinside=||,frame=single]
|\textbf{Link:}| |\url{http://alturl.com/8g2p9}|  
|\textbf{Project:}| Nova      
|\textbf{Type:}| Buffer overflow      
|\textbf{Developer:}| ...In a future patch that adds the ability to configure the executor type, we will need to deal with the issue you raise here.
\end{lstlisting}






We inspected the discussion and the follow-up submitted code to determine whether a security defect has been resolved. A security defect was considered resolved only when the situation meets the following three possible categories:
\begin{enumerate}
    \item Code is modified in the subsequent patchsets by the developer to resolve the security defect before the code change is merged. 
    \item Developer mentioned clearly in the reply to comments that the security defect has been fixed in another code change.
    \item The code change with the security defect was abandoned. As insecure code is not merged, it would not pose a harmful threat to the source code base. 
\end{enumerate}

\begin{figure*}[htbp]
  \centering
  \includegraphics[width=\linewidth]{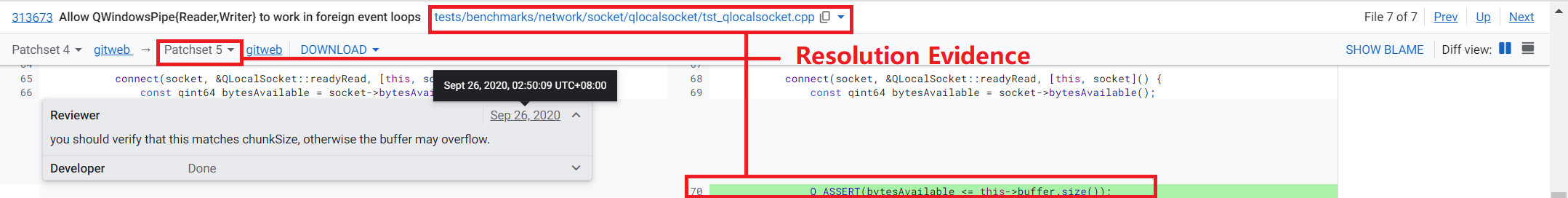}
  \caption{An example of adding an assert statement to fix buffer overflow.}
  \label{fig:resolution_evidence_no_RQ3}
\end{figure*}

As shown in Fig.~\ref{fig:resolution_evidence_no_RQ3}, the developer added an assert statement to check the buffer size in Line 70 of \verb|tst\_qlocalsocket.cpp| in patchset 5 to fix the buffer overflow, thus we can confirm that the security defect identified in this review comment was resolved. 

Employing the open coding and constant comparative method \cite{glaser1965constant}, we used MAXQDA\footnote{\url{https://www.maxqda.com/}} as the coding tool and extracted the solutions developers adopted from the specific code modification for fixing security defects in resolved instances, so as to investigate the common solutions of each security defect types.


\subsubsection{RQ3}
\noindent
To further understand why unresolved security defects were ultimately ignored by practitioners, we also utilized the open coding and constant comparative method~\cite{glaser1965constant} to examine the discussions between developers and reviewers. 

For the purpose of minimizing bias, this data extraction was performed by the first author and verified by two other co-authors. Any conflicts were discussed and addressed by the three authors, using a negotiated agreement approach~\cite{campbell2013coding}. The complete extraction results in this step is available online~\cite{replpack}.

\begin{table}
\scriptsize
  \caption{Data Items to be Extracted from Review Comments}
  \label{tab:Dataitems}
  \begin{tabular}{m{1.8cm}m{5.4cm}m{0.5cm}}
    \hline
    \textbf{Data Item}      &\textbf{Description}                                                                               &\textbf{RQ}            \\\hline
    Security-Related        & \textit{Whether the review comment is security-related.}                                          & RQ1                   \\\hline
    Security Defect         & \textit{The type of the identified security defect.}                                                & RQ1                   \\\hline
    Reviewer's Action       & \textit{The action suggested by the reviewer to cope with the security defect.}                                & RQ2.1                 \\\hline
    Developer's Action      & \textit{The action adopted by the developer to cope with the security defect.}                                 &{RQ2.2, RQ2.3}         \\\hline
    Solution                & \textit{The final coding solution adopted to fix the security defect.}	                            & RQ2.2                 \\\hline
    Relationship            & \textit{Relationship between the solution adopted by the developer and the solution suggested by the reviewer.}             & RQ2.3                 \\\hline
    Resolution Evidence     & \textit{The location of code modification for resolving the security defect.}                           & RQ2.4                   \\\hline
    Cause                   & \textit{The cause of not resolving the security defect.}                                                   & RQ3                   \\\hline
\end{tabular}
\end{table}

\section{Results}
\label{sec:result}

\subsection{RQ1: Category of Security Defects Identified in Code Reviews}
\noindent
As explained in Section \ref{sec:data_label}, 614 review comments were identified as security-related comments, which account for less than 1\% of all comments in code reviews. As detailed in Table~\ref{tab:security_related}, the majority of security defects (539 out of 614, 87.8\%) were identified by reviewers, which is considerably more than those raised by developers. Therefore this study is based on 539 security-related review comments that meet the former case. As described in Section \ref{sec:method}, we have predefined 15 types of security defects with their distribution (see Table~\ref{tab:type_frequency}). On the whole, we found that \textit{Race Condition} is the most frequently identified type and was discussed in as many as 39.0\% of instances. The second and third most frequently identified types are \textit{Crash} and \textit{Resource Leak}, accounting for 22.8\% and 10.9\%, respectively. There are 41 (7.6\%) review comments identified \textit{Integer Overflow}, followed by \textit{Improper Access} with 31 (5.8\%) instances. As can be seen in Table~\ref{tab:type_frequency}, there are also nine types that were identified on rare occasions with proportions lower than 5\%. Although \textit{SQL Injection} is a common network attack and listed as the top 10 web application security risks by the Open Web Application Security Project (OWASP) in the past 15 years~\cite{owasp2022topten}, no instance of this type was found in this study.
\begin{tcolorbox}[colback=white,colframe=black,left=0.05cm,right=0.05cm,top=0.05cm,bottom=0.05cm, sharp corners, boxrule=0.8pt]
\noindent
\textbf{RQ1 summary:} Security defects are not prevalently discussed in code review, with the proportion less than 1\%. Of those security-related review comments, a considerable amount of review comments detected the security defects \textit{race condition} (39.0\%), \textit{crash} (22.8\%), and \textit{resource leak} (10.9\%).
\end{tcolorbox}


\begin{table}
\scriptsize
  \caption{Frequency of each role that identified the security defects}
  \label{tab:security_related}
  \begin{tabular}{ |m{4.5cm}|m{1.5cm}|m{1.5cm}| } 
    \hline
    \textbf{Who Identified the Security Defect}   &\textbf{Number}    &\textbf{Percentage}   \\\hline
    Reviewer                                      & 539               & 87.8\%               \\\hline
    Developer                                     & 75                & 12.2\%               \\\hline
    \textbf{Total}                                & 614               & 100.0\%              \\\hline
\end{tabular}
\end{table}
\begin{table}
\scriptsize
  \caption{Frequency of each security defect types}
  \label{tab:type_frequency}
  \begin{tabular}{ |m{4.5cm}|m{1.5cm}|m{1.5cm}| } 
    \hline
    \textbf{Security Defect Type}   &\textbf{Number}    &\textbf{Percentage}    \\\hline
    Race Condition                  & 210               & 39.0\%                \\\hline
    Crash                           & 123               & 22.8\%                \\\hline
    Resource Leak                   & 59                & 10.9\%                \\\hline
    Integer Overflow                & 41                & 7.6\%                 \\\hline
    Improper Access                 & 31                & 5.8\%                 \\\hline
    Buffer Overflow                 & 24                & 4.5\%                 \\\hline
    Denial of Service (DoS)         & 12                & 2.2\%                 \\\hline
    Deadlock                        & 11                & 2.0\%                 \\\hline
    Encryption                      & 9                 & 1.7\%                 \\\hline
    Cross Site Script (XSS)         & 8                 & 1.5\%                 \\\hline
    Use After Free                  & 8                 & 1.5\%                 \\\hline
    Command Injection               & 1                 & 0.2\%                 \\\hline
    Cross Site Request Forgery      & 1                 & 0.2\%                 \\\hline
    Format String                   & 1                 & 0.2\%                 \\\hline
    SQL Injection                   & 0                 & 0.0\%                 \\\hline
    \textbf{Total}                  & 539               & 100.0\%               \\\hline
\end{tabular}
\end{table}

\subsection{RQ2: Treatment of Security Defects by Developers and Reviewers}

\noindent
\textbf{RQ2.1:} Table~\ref{tab:reviewer_actions} shows that over half of the reviewers (290 out of 539, 53.8\%) expected developers to \textit{fix} the identified security defects. A large portion (251 out of 290, 86.6\%) of these cases include specific solutions to assist developers in resolution, which may provide suggestions or even detailed code snippets for resolving the defects. Below is an example where the reviewer recommended that the developer should add verification logic to avoid the \textit{Buffer Overflow} defect.

\begin{lstlisting}[label={lst:sample},breaklines=true,breakatwhitespace=true,captionpos=b,basicstyle=\footnotesize\ttfamily,escapeinside=||,frame=single]
|\textbf{Link:}| |\url{http://alturl.com/tk8t9}|  
|\textbf{Project:}| Qt Base      
|\textbf{Type:}| Buffer Overflow      
|\textbf{Reviewer: you should verify that this matches}| 
        |\textbf{chunkSize}|,otherwise the buffer may overflow. 
\end{lstlisting}



Only 13.4\% (39 out of 290) of those fixes asserted that the defect needed to be fixed, without any guiding solutions. For example:

\begin{lstlisting}[label={lst:sample},breaklines=true,breakatwhitespace=true,captionpos=b,basicstyle=\footnotesize\ttfamily,escapeinside=||,frame=single]
|\textbf{Link:}| |\url{http://alturl.com/783po}|  
|\textbf{Project:}| Qt Base 
|\textbf{Type:}| Integer Overflow 
|\textbf{Reviewer:}| This could overflow too...|\textbf{You'll need to}|
       |\textbf{fix it}| otherwise the commit won't integrate. 
\end{lstlisting}



There are also 210 (39.0\%) cases where reviewers only identified security defects without indicating the next step that the developer should follow, which fall under the definition of \textit{Capture} type. Besides that, there were a few reviewers (39, 7.2\%) who explicitly suggested \textit{ignoring} the identified security defects for various reasons, such as the issues not worth fixing.

\noindent
\textbf{RQ2.2:} We inspected the discussion and subsequent patchsets to determine whether a security defect was fixed finally. As shown in Table~\ref{tab:developer_action}, developers chose to fix the identified security defect more often, accounting for 65.9\%. The actions developers took to each security issue are presented in Table~\ref{tab:developer_action_type}. The overall result is that almost every type of security defect has a fix rate upwards of 50.0\%, except for \textit{Deadlock}, as low as 36.4\%. As analyzed in RQ1, defects of type \textit{Race Condition}, \textit{Crash}, and \textit{Resource Leak} are the top three frequently identified security defects. As demonstrated in Table~\ref{tab:developer_action_type}, these types of security defect are also frequently addressed by developers in code reviews with fix rates of 64.3\%, 71.5\%, and 79.7\%, respectively. In addition to the top three types of security defects, there are another 11 security defect types from ``Integer Overflow'' to ``Format String'', totalling 147, and the fixing rate of these 11 types is comparatively low, at 57.8\% (85 out of 147).   

\begin{table}
\scriptsize
  \caption{Actions suggested by reviewers to cope with security defects}
  \label{tab:reviewer_actions}
  \begin{tabular}{ |m{4.5cm}|m{1.5cm}|m{1.5cm}| } 
    \hline
    \textbf{Reviewers' Actions}     &\textbf{Number}    &\textbf{Percentage}   \\\hline
    Fix with a Specific Solution    & 251               & 46.6\%               \\\hline
    Fix without a Specific Solution & 39                & 7.2\%                \\\hline
    Capture                         & 210               & 39.0\%               \\\hline
    Ignore                          & 39                & 7.2\%                \\\hline
    \textbf{Total}                  & 539               & 100.0\%              \\\hline
\end{tabular}
\end{table}
\begin{table}
\scriptsize
  \caption{Actions taken by developers to cope with security defects}
  \label{tab:developer_action}
  \begin{tabular}{ |m{4.5cm}|m{1.5cm}|m{1.5cm}| } 
    \hline
    \textbf{Developers' Actions}     &\textbf{Number}    &\textbf{Percentage}   \\\hline
    Resolve                          & 355               & 65.9\%               \\\hline
    Not Resolve                      & 161               & 29.9\%               \\\hline
    Unknown                          & 23                & 4.2\%                \\\hline
    \textbf{Total}                   & 539               & 100.0\%              \\\hline
\end{tabular}
\end{table}

\begin{table}
\scriptsize
  \caption{Developers' resolution rate for each security defect types}
  \label{tab:developer_action_type}
  \resizebox{\linewidth}{!}{
  \begin{tabular}{ |m{3cm}|m{1.5cm}|m{1.5cm}|m{1.5cm}| } 
    \hline
    \textbf{Security Defect Type}   &\textbf{Reviews}     &\textbf{Resolve}       &\textbf{Percentage}         \\\hline
    Race Condition                  & 210                 &135                & 64.3\%             \\\hline
    Crash                           & 123                 &88                 & 71.5\%             \\\hline
    Resource Leak                   & 59                  &47                 & 79.7\%             \\\hline
    Integer Overflow                & 41                  &26                 & 63.4\%             \\\hline
    Improper Access                 & 31                  &17                 & 54.8\%             \\\hline
    Buffer Overflow                 & 24                  &14                 & 58.3\%             \\\hline
    Denial of Service (DoS)         & 12                  &6                  & 50.0\%             \\\hline
    Deadlock                        & 11                  &4                  & 36.4\%              \\\hline
    Encryption                      & 9                   &5                  & 55.6\%             \\\hline
    Cross Site Script (XSS)         & 8                   &4                  & 50.0\%             \\\hline
    Use After Free                  & 8                   &6                  & 75.0\%             \\\hline
    Command Injection               & 1                   &1                  & 100.0\%            \\\hline
    Cross Site Request Forgery      & 1                   &1                  & 100.0\%            \\\hline
    Format String                   & 1                   &1                  & 100.0\%            \\\hline
    \textbf{Total}                  & 539                 &355                & 65.9\%             \\\hline
\end{tabular}
}
\end{table}

\begin{figure}[htbp]
  \centering
  \includegraphics[width=\linewidth]{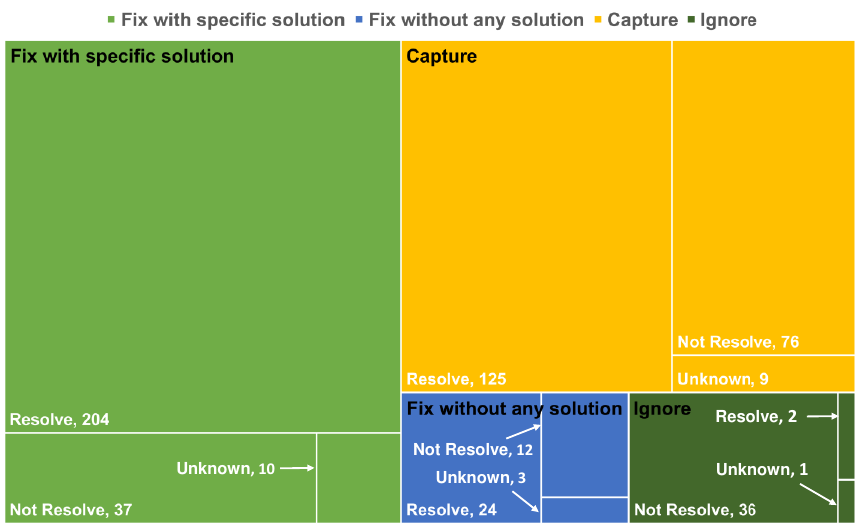}
  \caption{The treemap of relationship between the action developers taken and the action reviewers suggested}
  \label{fig:relationship_update_action}   
\end{figure}

\noindent
\textbf{RQ2.3:} The relationship between the action developers took and the action reviewers suggested is illustrated in Fig.~\ref{fig:relationship_update_action}. When reviewers provide a clear idea for fixing the identified security defects with specific solutions (\textit{Fix with a specific solution}), the fix rate by developers reaches 81.3\% (204 out of 251). When reviewers only point out that fixing is needed but do not offer any guidance (\textit{Fix without a specific solution}), developers choose to address these defects in 61.5\% (24 out of 39) of the cases. Furthermore, when reviewers indicate the existence of security defects without further instructions (\textit{Capture}), only 59.5\% (125 out of 210) of these issues are fixed. Based on our findings, it can be speculated that reviewers' suggestions that include guidance in code review, such as whether and how to resolve defects, are crucial to improving the overall fix rate of identified security defects. 
As shown in Fig.~\ref{fig:relationship_update_action}, for the instances in which the actions suggested by reviewers are the \textit{Fix} type, 78.6\% (228 out of 290) of developers fixed the identified security defects. For the instances in which reviewers suggested ignoring the defects, nearly all (36 out of 39, 92.3\%) developers ignored the defects. Overall, it can be concluded that the majority of developers tend to agree with reviewers' opinions when the reviewers express clear perspectives on defect handling. Hence, the participation and enthusiasm of reviewers are crucial for detecting security defects during code review.

\noindent
\textbf{RQ2.4}: The coding solutions adopted by developers to resolve different security defects are further investigated and presented in Table~\ref{tab:common_solutions}. In order to make the sample size large enough to ensure the credibility of the conclusions, we only selected the top three security defects based on their prevalence for analysis, i.e., \textit{Race Conditions}, \textit{Crash}, and \textit{Resource Leak}. 

In terms of \textit{Race Condition}, the most common approach adopted by developers is to \textbf{take thread-safety measures}. These measures include using thread-safe functions, such as atomic operations, the \verb|invokeMethod| function, or synchronization functions that utilize signals and slots (e.g., \verb|QFutureWatcher| in Qt). 
They also employed custom logic when working with resources, including measures such as adding locks, usage limitations, and updating before usage to ensure consistency. \textbf{Code refactoring} is also an important solution for \textit{Race Condition}, with 33 instances. A few cases adopted \textbf{concurrency management}, which includes passing messages between threads and adding wait functions. Additionally, 7 developers solved the issue by \textbf{handling side effects}, which means dealing with the consequences of \textit{Race Conditions} indirectly, such as capturing exceptions.

In the instances of \textit{Crash}, there are five possible solutions. \textbf{Code refactoring} and \textbf{adding condition check} are the two main solutions adopted by developers to fix the \textit{Crash} defects. In 13 review comments, developers \textbf{captured exceptions by try/catch block} to avoid \textit{Crash}. Furthermore, 6 developers \textbf{safely terminated execution in advance} to prevent damage caused by an abrupt \textit{Crash}, and a specific example of this case is to add an assert statement to immediately trigger an exception and terminate the execution of the program, if a certain condition or constraint is not met. There are also 4 cases where developers \textbf{used safe functions} that can eliminate potential exceptions, thus improve the overall stability of the program and minimize the likelihood of crashing.

Approximately half of \textit{Resource Leak} defects are fixed by \textbf{adding resource release functions}, where developers may explicitly close resources or prevent skipping of the deletion function through modification in code logic. 9 developers also \textbf{used resource-management techniques}, such as smart pointers, Resource Acquisition Is Initialization (RAII), or bridge technologies during fixing. Additionally, 8 developers \textbf{reduced resource allocation} to avoid leaks through converting to passing by reference, transferring resource ownership and so on. Only 4 cases involve \textbf{code refactoring} as a solution, while just 2 cases addressed the security defects through \textbf{handling side effects}, as previously mentioned.

\begin{table}
\scriptsize
  \caption{The solutions of the top three most frequent security defect types}
  \label{tab:common_solutions}
  \begin{tabular}{ m{2cm}m{4.5cm}m{1cm} } 
    \hline
    \textbf{Type}                       &\textbf{Solution}                            &\textbf{Number}                                  \\\hline          
    \multirow{4}*{Race Condition}       & {Take thread-safety measures}               & 55                                              \\
    ~                                   & {Code refactoring}                          & 33                                               \\
    ~                                   & {Concurrency management}                    & 9                                                \\
    ~                                   & {Handle side effects}                       & 7                                                \\\hline
    \multirow{5}*{Crash}                & {Code refactoring}                          & 30                                                \\
    ~                                   & {Add condition check to avoid crash}        & 26                                                \\
    ~                                   & {Capture exceptions by try/catch block}     & 13                                                \\
    ~                                   & {Safely terminate execution in advance}     & 6                                                \\
    ~                                   & {Use safe functions}                        & 4                                                 \\\hline
    \multirow{5}*{Resource Leak}        & {Add release function}                      & 20                                                \\
    ~                                   & {Use resource-management techniques}        & 9                                                  \\
    ~                                   & {Reduce usage of resource}                  & 8                                                 \\
    ~                                   & {Code refactoring}                          & 4                                                  \\
    ~                                   & {Handle side effect}                        & 2                                                   \\\hline
\end{tabular}
\end{table}

\begin{tcolorbox}[colback=white,colframe=black,left=0.05cm,right=0.05cm,top=0.05cm,bottom=0.05cm, sharp corners, boxrule=0.8pt]
\noindent
\textbf{RQ2 summary:} 53.8\% of the reviewers indicated a need to fix the identified security defects after their detection, and most of them were willing to provide specific solutions for developers to fix the defects. From the developers' perspective, majority of developers tend to agree with reviewers' suggestions, and over half of the identified security defects were resolved by developers. 
\end{tcolorbox}

\noindent
\subsection{RQ3: Causes of Not Resolving Security Defects}

\noindent
According to the aforementioned result of RQ2.2, there are 161 instances where the identified security defects were not resolved. By manually inspecting the discussion for each review comments, we excluded 64 (39.8\%) review comments neither developers nor reviewers involved in these instances clearly indicate the causes for ignoring the identified security defects, leaving us with 97 instances (60.2\%) for further analysis. The statistical results of the remaining instances can be found in Table~\ref{tab:cause}, and six causes were then identified.

Nearly half of (44, 45.4\%) unresolved security defects are because either developers or reviewers think it is \textit{Not worth fixing the defect now}, which is the most common cause of not resolving the identified security defects. From the perspective of security defects, the identified security defects in these cases may be harmless and acceptable for developers, or the occurrence scenarios of security defects are so tricky that they will not become system hazards under normal utilization. It may also be because that there are other security defects in the code that will have a greater impact, and those currently found are negligible comparatively. On the developer side, fixes might cost too much effort and require tons of changes. 
If existing solutions had other adverse effects on the system and were irreconcilable, developers would also choose to ignore the identified defects in light of the benefit of current code changes. In addition, some developers noted that the resolutions for identified security defects were not an immediate concern and could be considered in the future. Two examples corresponding to the above two situations are presented below, the cruxes has been emphasized in bold:
\begin{lstlisting}[label={lst:sample},breaklines=true,breakatwhitespace=true,captionpos=b,basicstyle=\footnotesize\ttfamily,escapeinside=||,frame=single]
|\textbf{Link:}| |\url{http://alturl.com/hm4o7}|  
|\textbf{Project:}| Nova 
|\textbf{Developer:}| after discussing it on IRC [1], we went on a consensus that it's |\textbf{acceptable}| to remove the VIF from the metadata since the NIC on the VM already detached, even if the Neutron action could potentially fail.
\end{lstlisting}





\begin{lstlisting}[label={lst:sample},breaklines=true,breakatwhitespace=true,captionpos=b,basicstyle=\footnotesize\ttfamily,escapeinside=||,frame=single]
|\textbf{Link:}| |\url{http://alturl.com/xcb6n}|  
|\textbf{Project:}| Qt Base
|\textbf{Developer:}| I'll leave it as it is - otherwise I'll
    have to |\textbf{change the example too much and write}|
    |\textbf{tons of code}| obscuring the real example.
\end{lstlisting}



We attribute \textit{Disagreement between the developer and the reviewer} to be the reason why developers do not resolve the security defects in 33 review comments (34.0\%). In these cases, some developers could not comprehensively understand reviewers' opinions, while others indicated that the identified security defects did not exist. Furthermore, some developers believed that fixing was unnecessary or the solution was unreasonable. In the following example, the developer objected to the reviewer's suggestion to control traffic by adding a security group, asserting that no modification was required.
\begin{lstlisting}[label={lst:sample},breaklines=true,breakatwhitespace=true,captionpos=b,basicstyle=\footnotesize\ttfamily,escapeinside=||,frame=single]
|\textbf{Link:}| |\url{http://alturl.com/9uwgv}|  
|\textbf{Project:}| Neutron
|\textbf{Reviewer:}| i suspect this requires security groups.
|\textbf{Developer:}| Why we need this? Could you explain because I don't think we need anything here.
\end{lstlisting}




Due to the lack of knowledge or limitation by other system logic, 11.3\% of identified security defects were ignored for the reason that practitioners \textit{had no effective solution to thoroughly resolve the defects}, and below is an example of this case:
\begin{lstlisting}[label={lst:sample},breaklines=true,breakatwhitespace=true,captionpos=b,basicstyle=\footnotesize\ttfamily,escapeinside=||,frame=single]
|\textbf{Link:}| |\url{http://alturl.com/t5paz}|  
|\textbf{Project:}| Qt Creator
|\textbf{Reviewer:}| ...If you are not happy with a crash you can add a check against 0. This will avoid the crash here but I am pretty sure that it will crash sooner or later on a different location...
\end{lstlisting}



In 6.2\% review comments, the reason for not resolving security defects is that the resolution is considered \textit{out of the scope of the commit}. As shown in the example below, the identified security defect was historical and thus orthogonal with the feature of this commit. Accordingly, the developers reckoned that those defects should wait to be resolved in specific logic changes in the future, rather than now.
\begin{lstlisting}[label={lst:sample},breaklines=true,breakatwhitespace=true,captionpos=b,basicstyle=\footnotesize\ttfamily,escapeinside=||,frame=single]
|\textbf{Link:}| |\url{http://alturl.com/joasw}|  
|\textbf{Project:}| Nova
|\textbf{Developer:}| ...I think that the multi-attach problem is |\textbf{orthogonal}| and should be investigated in another patch.
\end{lstlisting}




In addition, three occasional instances were found, in two (2.1\%) of which the developers believed that it was \textit{users' responsibility to make correct choices} to guarantee the system running appropriately, and no any modification was conducted to the source code. While in the remaining one (1.0\%), the developer clearly indicated that he/she \textit{had no time to rework} and left the reviewer to accept identified defects or directly abandon the whole change. 

\begin{tcolorbox}[colback=white,colframe=black,left=0.05cm,right=0.05cm,top=0.05cm,bottom=0.05cm, sharp corners, boxrule=0.8pt]
\noindent
\textbf{RQ3 summary:} Generally speaking, 39.8\% related instances did not provide the cause of failure to resolve. \textit{Not worth fixing the defect now} and \textit{Disagreement between the developer and the reviewer} are the main reasons of ignoring security defects.
\end{tcolorbox}

\begin{table}
\scriptsize
  \caption{The distribution of causes for ignoring security defects identified in code reviews}
  \label{tab:cause}
  \begin{tabular}{ |m{5.5cm}|m{1cm}|m{1cm}| }
    \hline
    \textbf{Cause}                                      &\textbf{Number}       &\textbf{\%}             \\\hline
    Not worth fixing the defect now                     & 44                   & 45.4\%                 \\\hline
    Disagreement between the developer and the reviewer & 33                   & 34.0\%                 \\\hline
    No clear solution to solve the problem              & 11                   & 11.3\%                 \\\hline
    Out of the scope of this commit                     & 6                    & 6.2\%                  \\\hline
    Think users should make correct choices             & 2                    & 2.1\%                  \\\hline
    Developers had no time to rework                    & 1                    & 1.0\%                  \\\hline
    \textbf{Total}                                      & 97                   & 100.0\%                \\\hline
\end{tabular}
\end{table}

\section{Implications}
\label{sec:implications}
Here we discuss several implications of the findings reported in this paper.

\noindent
\textbf{A two-step of detection mechanism is suggested to conduct security practices in software development.} Our study found that in the process of code review, the majority of reviewers provided useful suggestions to fix identified security defects, and developers usually agreed and adopted solutions suggested by reviewers. This indicates that reviewers' assessment of security defects is trustworthy for developers. Generally speaking, code review is effective in detecting and addressing security defects. Although various tools (e.g., SAST, DAST, IAST) have been used in modern code review to speed up the review process, these tools test only based on known scenarios and have limitations in test coverage, thus resulting in potential false positives~\cite{singh2020automated}. Experienced and knowledgeable code reviewers, due to their deeper understanding of code context, can capture security defects which do not conform to known patterns and cannot be detected by tools. Therefore, automated tools and code review, as two significant approaches of security defect detection, need to complement each other. We recommend a two-step detection mechanism that combines the two approaches: tools to conduct scalable and fast security defect detection as the first check, and then the reviewers to conduct code review referring to the detection results of the tool. During the second step, the reviewers check the results generated by tools to provide developers with instructions for further action, and at the same time review the submitted code to find defects that the tool failed to detect. This mechanism not only improves efficiency, but also enhances the comprehensiveness of security defect detection.

\noindent
\textbf{The characteristics of the project can affect the type and quantity of security defects found in it.} We found that \textit{XSS (Cross-Site Scripting)} and \textit{SQL Injection} are less discussed during code review, which is consistent with the findings of Paul \textit{et al.}~\cite{paul2021security}, but contrary to the results of di Biase \textit{et al.}~\cite{di2016security}, which demonstrated that XSS was a frequently identified security defect with a relatively higher number than other types. The projects used in this study (Nova, Neutron, Qt Base, and Qt Creator) and Paul \textit{et al.}'s work (Chromium OS) are the projects that provide infrastructure for higher-level applications to run on, with less direct interaction with users' inputs and outputs, while di Biase \textit{et al.} selected Chromium, a Web browser that has multiple ways of directly interacting with users. One possible reason for this result is that the likelihood of potential input/output-related security defects in core components and projects may be low. This further confirms that project characteristics can influence the types and quantity of security defects that may exist in this project. 

\noindent
\textbf{Reviewers need to pay more attention to high-risk code with the use of multi-threading or memory allocation.} \textit{Race Condition} and \textit{Resource Leak} related security defects are frequently identified in code review. These two defect types are also widely recognized as common defects in software development~\cite{wu2016relda2, zhang2017rclassify}. Hence, we encourage code reviewers to conduct a rigorous inspection of code involving multi-threading and memory allocation during code review, as they can potentially introduce \textit{Race Condition} and \textit{Resource Leak} defects, making them more susceptible to security risks.\\
\noindent
\textbf{Appropriate standardization of practitioners' behaviors in code review is critical for better detection of security defects.} In modern code review, strict reviewing criteria are not mandated~\cite{shihab2011high}. We found that some developers' and reviewers' actions result in ambiguity during the code review process. For example, some comments that identified security defects were neither responded nor had corresponding code modifications. Hence, code reviews may not foster a sufficient amount of discussion~\cite{mcintosh2016empirical}, increasing the time and effort of the development process and having a negative impact on software quality. Here are several specific recommendations regarding standardization:
(1) For security defects that remain unresolved due to disagreement between developers and reviewers, reviewers could further assess the risk of the security defects. We found that the main reason for not resolving security defects is \textit{Disagreements between the developer and the reviewer} in which the developer did not agree with the reviewer's assessment, and thus decided not to fix the security defects identified. However, due to the different knowledge and experience, it is likely for the developer to merge risky security defects into the source code. Hence, we suggest that when there is a disagreement, reviewers should further assess the risk of identified security defects and communicate with developers if necessary.
(2) It is preferable for developers to resolve identified security defects. However, when developers decide not to address a security defect (possibly due to risk assessment or cost-benefit considerations), they should provide clear reasons for this decision in the discussion. It was found that in 40\% of the cases, the identified security defects were left unresolved, with no reasons provided. This negatively impacts adequate communication between reviewers and developers, making review details opaque and untraceable. Therefore, we recommend that when a security defect was decided to be left unresolved, sufficient justifications should be provided in the discussion to facilitate further handling of the unresolved security defects. 
(3) Unresolved security defects should be properly documented, and the developers who decide to fix them in the future should be clearly scheduled for resolution in subsequent stages. According to the results of RQ2.2, 29.9\% of security defects were unresolved and merged into source code
Documenting unresolved security defects in code review helps to effectively track and manage them. Clearly scheduling unresolved security defects that developers decide to fix in the future can ensure they are actually resolved in a timely manner, thus preventing them from causing damage to the system. Therefore, we encourage practitioners to document unresolved defects and schedule needed fixes.

\section{Threats to Validity}
\label{sec:threat}

\textbf{Internal Validity:} During the data processing phase, there are comments that were either generated by bots or related to non-review target files, which could influence the accuracy of the final results. We filtered these comments 
to mitigate bias. Furthermore, we employed a keyword-based search approach to obtain potential security-related comments, which can lead to missing security-related comments that do not contain the exact keywords. 
To reduce this bias, we collected all the keywords utilized in previous studies into the keyword list and refined the list according to the approach proposed by Bosu \textit{et al.}~\cite{bosu2014identifying}, ensuring a comprehensive set of keywords to cover all eligible review comments as much as possible. 


\textbf{External Validity:} We selected four projects from the OpenStack and Qt communities (two each) as the primary data source of our study. However, these projects may not fully represent the entire landscape of security defects across all software systems. This limitation poses a potential threat to the generalizability of our results. To address this concern, we compared and discussed with the previous studies that explored similar questions to supplement our own findings and reduce the risk of interpretation bias. 



\textbf{Construct Validity:} Since this study predefined the types of security defects and matched practical scenarios with security defect types through manual inspection, there is a potential cognitive bias arising from subjective judgments. To reduce this bias, we based the classification on the security defect types proposed in previous works~\cite{paul2021security} and clarified these security defects by CWEs, thus ensuring the concepts of each type are accurate, appropriate, and consistent throughout the entire research process.
In addition, all the data labeling and extraction processes in this study were carried out manually, which introduces the possibility of subjective and potentially misleading conclusions. Therefore, during the data labelling phase, the first and second authors conducted a pilot data labelling independently and reached a consensus on labelling criteria through discussions. During the data extraction phase, while the first author performed the extraction work, the second and third authors reviewed the results to ensure the accuracy and comprehensiveness of the data extraction results.



\textbf{Reliability:} We drafted a protocol outlining the detailed procedure before conducting our study. The protocol was reviewed and confirmed by all authors to ensure the clarity and repeatability of the method.
We also made our full dataset available online for future replications \cite{replpack}. 

\section{Conclusions}
\label{sec:conclusions}
In this work, we investigated the security defects identified in code review comments. We analyzed the data from four open source projects of two large communities (OpenStack and Qt) that are known for their well-established code review practices. More specifically, we manually inspected 20,995 review comments obtained by keyword-based search and identified 614 security-related comments. We extracted the following data items from each comment: 1) the type of security defect, 2) the action taken by reviewers and developers, 3) reasons for not resolving identified defects from these comments. Our main results are: (1) security defects are not widely discussed in code reviews, and when discussed, \textit{Race Condition} and \textit{Crash} security defects are the most frequently identified types; (2) the majority of the reviewers express explicit fixing suggestions of the detected security defects and provide specific solutions. Most of the developers are willing to agree with reviewers' opinions and adopt their proposed solutions; 
(3) \textit{Not worth fixing the defect now} and \textit{Disagreement between the developer and the reviewer} are the main reasons for not resolving security defects.




%





\bibliographystyle{ieeetr}
\bibliography{reference}


\balance

\end{document}